\documentclass[
 reprint,
superscriptaddress,
 amsmath,amssymb,
 aps,
prstab,
]{revtex4-2}

\usepackage{graphicx}
\usepackage{dcolumn}
\usepackage{bm}
\usepackage{dblfloatfix}
\usepackage{siunitx}

\usepackage{hyperref}
\hypersetup{
    breaklinks=true,   
    colorlinks=true,   
    allcolors=black,
        }

\graphicspath{{Images/}}

\begin{document}

\preprint{APS/123-QED}

\title{Wakefield damping in a distributed coupling LINAC}

\author{Evan Ericson}
\affiliation{
 CERN, European Organization for Nuclear Research, Geneva 1221, Switzerland 
}
\affiliation{
 Department of Physics and Engineering Physics, University of Saskatchewan, Saskatoon, Canada.
}
\author{Alexej Grudiev}\email{alexej.grudiev@cern.ch}
\affiliation{
 CERN, European Organization for Nuclear Research, Geneva 1221, Switzerland 
}
\author{Drew Bertwistle}
\affiliation{
 Department of Physics and Engineering Physics, University of Saskatchewan, Saskatoon, Canada.
}
\affiliation{%
 Canadian Light Source, Saskatoon, Canada
}
\author{Mark J. Boland}
\affiliation{
 Department of Physics and Engineering Physics, University of Saskatchewan, Saskatoon, Canada.
}
\affiliation{%
 Canadian Light Source, Saskatoon, Canada
}

\date{\today}

\begin{abstract}
The number of cells in a $\pi$-mode standing wave (SW) accelerating structure for the Compact linear Collider (CLIC) project is limited by mode overlap with nearby modes. The distributed coupling scheme avoids mode overlap by treating each cell as independent. Designs of cells suitable for distributed coupling with strong wakefield damping have not previously been studied. In this paper we develop a SW cell to be used in a distributed coupling structure that can satisfy the CLIC transverse wakepotential limit. From the middle cell of the CLIC-G* travelling wave (TW) structure, a SW cell is designed. The cell is adapted to be suitable for distributed coupling. Its wakepotentials in an ideal case of open boundaries are reduced to satisfy the wakepotential threshold. An electric boundary is added to the model to simulate total reflection at the distribution network. A horizontal coupler cell that connects to the distribution network such that the reflected wakefields remain similar to the open boundary case is simulated. A triplet module which takes advantage of cell-to-cell coupling to reduce reflected wakepotential is presented.
\end{abstract}

\maketitle

\section{\label{sec:intro}Introduction}
\begin{figure}
    \includegraphics[width=5cm]{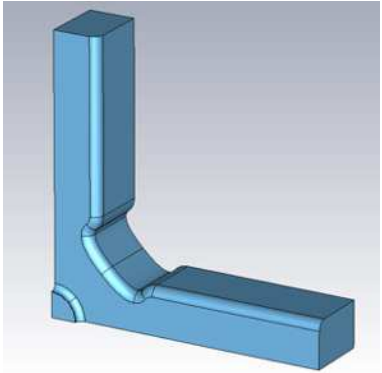}
    \caption{Accelerating cell from the CLIC-G* TW structure for CLIC with quarter symmetry.}
    \label{fig:twCellGeometry}
\end{figure}

The main linear accelerator (LINAC) of the CLIC project  uses accelerating structures to generate high-energy particles for particle physics experiments \cite{Aicheler2012}. The current LINAC design consists of TW X-band radio frequency (RF) accelerating structures. These structures consist of 26 regular accelerating cells and two coupler cells. The cell irises radii are tapered from \SI{3.15}{\milli\meter} to \SI{2.35}{\milli\meter} to maintain a constant loaded gradient of \SI{100}{\mega\volt/\meter}. All cells feature waveguides with silicon carbide loads to dissipate power from higher order modes (HOM) \cite{Zha2016}. Figure \ref{fig:twCellGeometry} shows an example of a TW cell.

Recent work has shown converting the $2\pi/3$-mode TW structure to a $\pi$-mode SW structure of equal length gives an increase in rf-to-beam efficiency of $\sim$ \SI{3}{\percent} \cite{VasimKhan2013}. A SW structure operating in the $\pi$-mode is limited to the number of cells, N, given by Equation \ref{eq:nCells}.

\begin{equation}
 \sqrt{\frac{Q_{\pi}\pi^{2}k}{4}} > N.
 \label{eq:nCells}
\end{equation}
For a SW cell based on CLIC-G* cell shape, the maximum number of cells before mode overlap is 9. With this many cells, the $\pi$-mode is separated from the next adjacent mode by

\begin{equation}
    \Delta f^{\pi}_{N,N-1} = f_{r}\frac{k\pi^{2}}{4N^{2}} = \SI{2}{\mega\hertz}.
\end{equation}
Operating a side-coupled structure in the $\pi/2$-mode simultaneously increases the number of cells in a structure before mode overlap while keeping the shunt impedance high. A side-coupled geometry is not easily amenable to cells with four HOM waveguides.  Distributed coupling topologies also allow for a SW structure to increase the number of cells by treating each cell as independent. In this configuration, each accelerating cell is connected directly to the power source by a waveguide which runs parallel to the structure \cite{Tantawi2020,Jiang2021}. Because the structure does not rely on coupling between cells, the cells' iris aperture can be made small to increase the shunt impedance. Distributed coupling schemes can also increase high-gradient performance by limiting the amount of power flow through the cell irises which are prone to RF breakdown events \cite{Dolgashev2010,Simakov2018}. As the iris aperture decreases, the effects of wakefields increase \cite{Bane2003}. Existing designs reduce wakefields by use of detuning such that problematic dipole modes add coherently later, ideally after the end of the bunch train \cite{Rimmer1993}. Detuning can provide dipole modes with Q$_{ext} = 1000$ while CLIC requires modes with Q$_{ext} \sim 10$ \cite{Bane2018a,Grudiev2010}. In this work, we present the development of an RF accelerating cell suitable for a distributed coupling structure that satisfies the CLIC transverse wakefield damping requirement. Section \ref{sec:openWakes} outlines the procedure followed to get a SW cell design. The cell is adapted to include a power input and then further to reduce the wakepotential of the cell. In Section \ref{sec:reflectedWakes} the reflected wakepotential from a distribution network is analyzed and Section \ref{sec:reflectedDesigns} describes cell designs capable of reducing the reflected wakepotentials to an acceptable level. The final findings are summarized in \ref{sec:conc}.

\section{\label{sec:openWakes}Open boundary wakes}

The middle cell of the TW structure was chosen as a starting point for the design. The cell length was adjusted so the $\pi$-mode was synchronous with a relativistic beam. Parameters sweeps of the iris thickness and ellipticity were performed in CST's eigenmode solver \cite{CST} to select an iris shape that minimized the surface electric field. Sweeps of the outer wall parameters A0 and A2, described in \cite{Zha2016}, confirmed the values of the TW cell also minimized the surface magnetic field in this case. The outer wall was adjusted to obtain the operating frequency. The RF parameters of the resulting cell are listed in Table \ref{tab:swCell}. An accelerating structure based on this SW cell design having the same length as the TW structure would have an RF-to-beam efficiency of \SI{31}{\percent}.

\begin{table}[h]
\caption{RF parameters of SW cell.}
\begin{ruledtabular}
\begin{tabular}{ccc}
\textbf{Quantity} & \textbf{Unit} & \textbf{Value}\\
\colrule
cell outer radius, b & \SI{}{\milli\meter} & 8.5638 \\
iris thickness, d & \SI{}{\milli\meter} & 2.6 \\
iris ellipticity, e & \SI{}{\milli\meter} & 2.4 \\
f$_{0}$ & \SI{}{\giga\hertz} & 11.9945 \\
Q$_{0}$ & - & 6,927 \\
R/Q$_{0}$ & \SI{}{\ohm/\meter} & 10,962 \\
R & \SI{}{\mega\ohm} & 75.93 \\
E$_{s}$/E$_{a}$ & - & 1.80 \\
H$_{s}$/E$_{a}$ & \SI{}{\milli\ampere/\volt} & 4.67 \\
S$_{c}$/E$_{a}^{2}$ & \SI{}{\milli\ampere/\volt} & 0.35
\end{tabular}
\end{ruledtabular}
\label{tab:swCell}
\end{table}

\begin{figure}[h]
    \includegraphics[width=3cm]{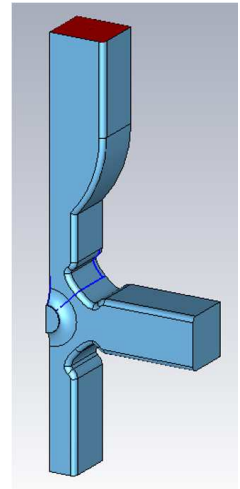}
    \caption{Long slot cell geometry with half symmetry. The cell is matched by adjusting the cell outer radius and the length of the HOM waveguide between the cell and the coupler waveguide.}
    \label{fig:longSlotGeometry}
\end{figure}

A cell with four HOM waveguides is fed by cell-to-cell coupling. To make the cell suitable for a distributed coupling structure, a coupler cell was generated based on the TW coupling cell. The TW coupler cells have only two HOM waveguides and therefore the expected wakepotential of a structure of repeating coupler cells was higher than the four HOM waveguide cell. The SW coupler cell included only one coupler waveguide and three HOM waveguides. The cell was matched to the source by adjusting the cell's outer radius and the aperture between the coupler waveguide and the cell. The x-polarization of the wakepotential was found to be lower in amplitude than the y-polarization. The evaluated transverse wakepotential of the coupler cell wacs above the threshold of \SI{6}{\volt(\pico\coulomb\cdot\meter\cdot\milli\meter)^{-1}} \cite{Schulte2010}. A hybrid cell design with a HOM waveguide transition between the coupler waveguide and the cell shown in Figure \ref{fig:longSlotGeometry} had a transverse wakepotential between the coupler cell and four HOM waveguide cell at the second bunch position.

\begin{figure}[h]
    \includegraphics[width=8cm]{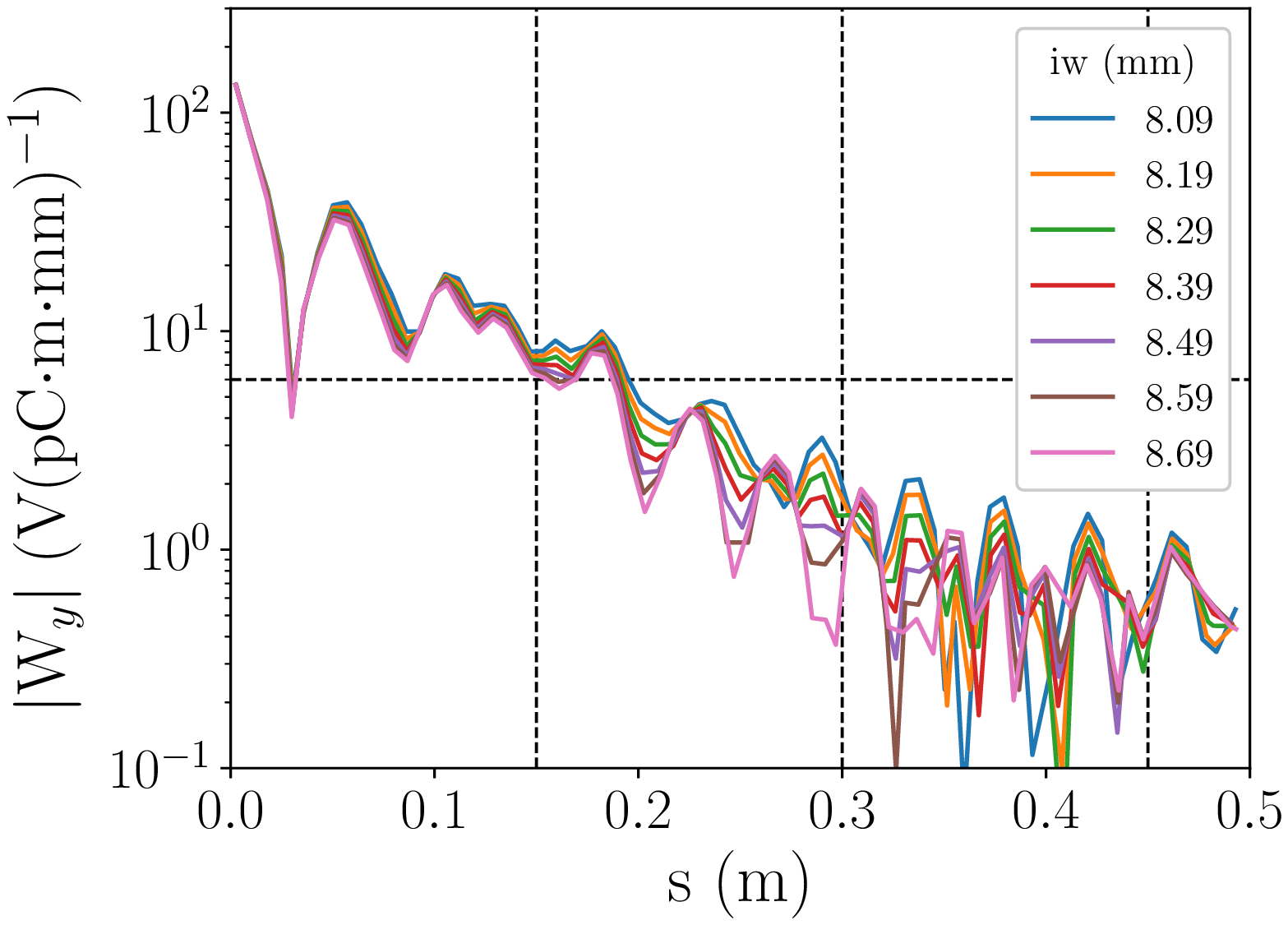}
    \includegraphics[width=8cm]{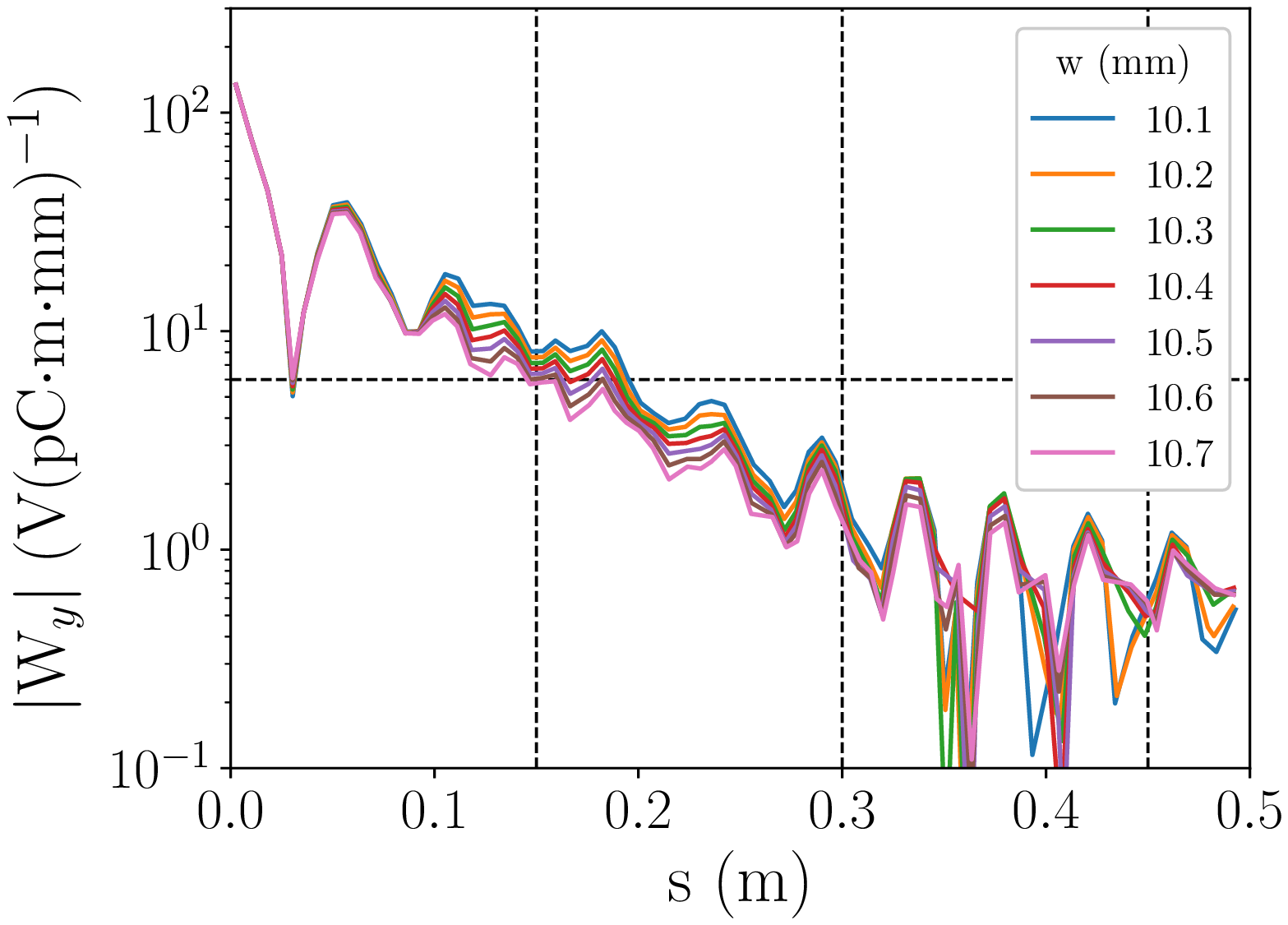}
    \caption{Transverse wakepotential envelopes of long slot cell with various HOM aperture widths, iw (top) and various HOM waveguide widths, w (bottom).}
    \label{fig:longSlotSweeps}
\end{figure}

\begin{figure}[h]
    \includegraphics[width=8cm]{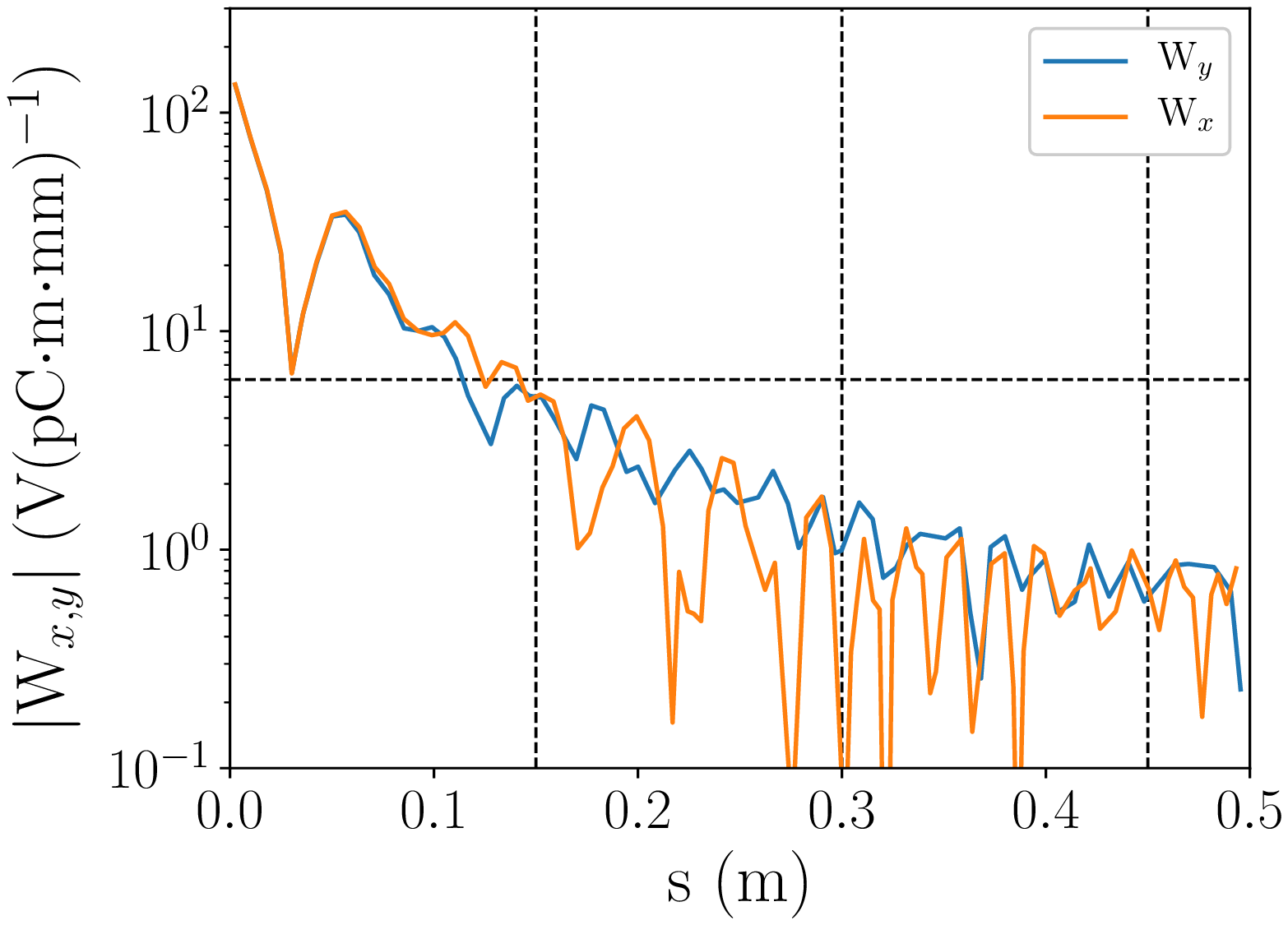}
    \includegraphics[width=8cm]{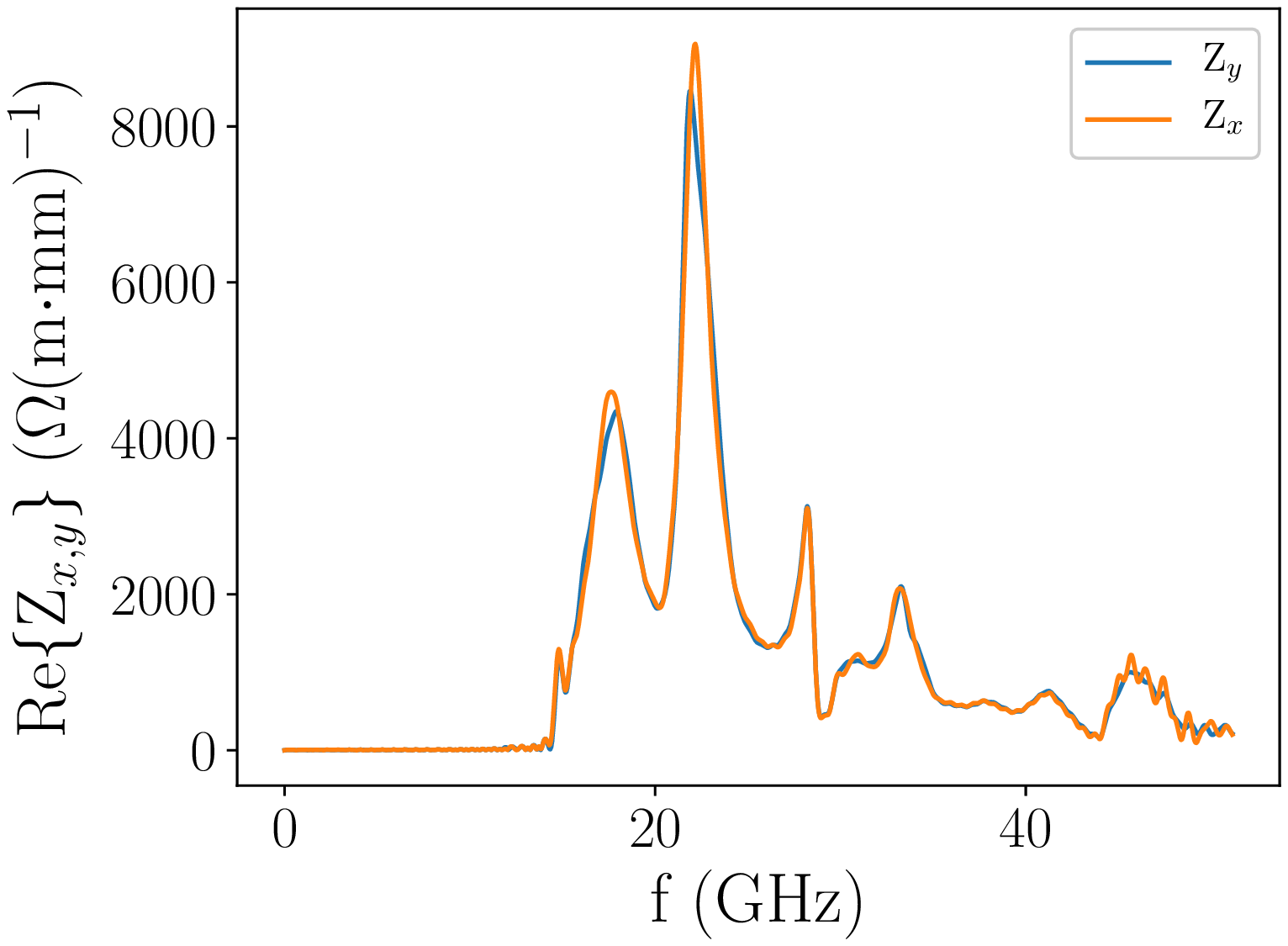}
    \caption{Transverse wakepotentials (top) and impedances (bottom) of long slot cell w = \SI{10.7}{\milli\meter} (bottom).}
    \label{fig:longSlotWakeImp}
\end{figure}

To decrease the wakepotential below the threshold, two approaches were taken. First, the aperture of the HOM waveguides were increased. Changing the geometry in this way caused the Q$_{ext}$ of the dominant dipole mode to decrease at the expense of a lower Q$_{0}$ and a larger expected temperature rise on the surface of the cell. The cells needed to be tuned for each simulated value. The second approach was to increase the HOM waveguide widths. The wakepotentials from these sweeps are shown in Figure \ref{fig:longSlotSweeps}. The wakepotential at the second bunch location was more sensitive to changes in w. A value of w = \SI{10.7}{\milli\meter} was selected. Both polarization of the transverse wakepotentials and their impedances are shown in Figure \ref{fig:longSlotWakeImp}.

\section{\label{sec:reflectedWakes}Reflected wakes}

\begin{figure}[h]
    \includegraphics[width=4cm]{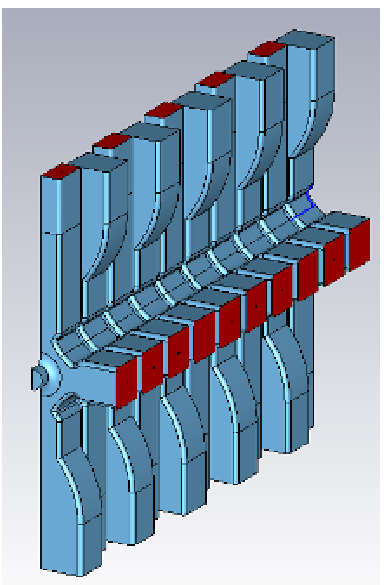}
    \caption{Model simulated to obtain wakepotential of reflected wakefields of the long slot cell using half symmetry. The HOM waveguide ports (red) are connected to an open boundary. The coupler waveguides are slightly shorter than the HOM waveguides and experience an electric boundary. The model was parameterized to maintain these boundary conditions as the coupler waveguide length was increased.}
    \label{fig:longSlotStructure}
\end{figure}

To model the effect of a distribution network, parallel waveguide or otherwise, on the wakepotential of the structure, an electric boundary condition was placed on the coupler waveguide surfaces. This represents a worst-case-scenario in which all the wakefields extracted through the coupler waveguide are reflected back into the cell where they may act on following bunches. Figure \ref{fig:longSlotStructure} shows the geometry simulated to include the effect of perfect reflection at the coupler waveguide.

\begin{figure}[h]
    \includegraphics[width=8cm]{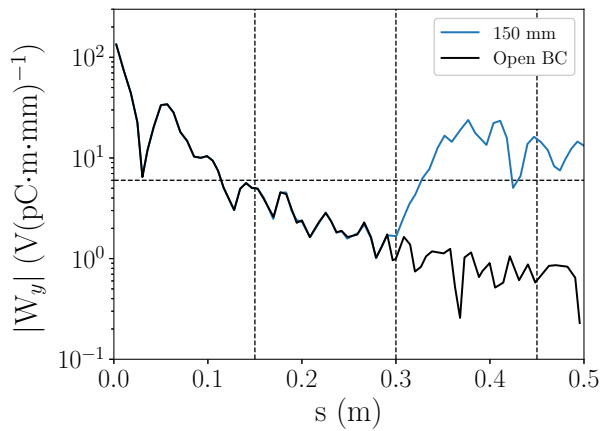}
    \caption{Reflected transverse wakepotential envelopes of long slot cell with w = \SI{10.7}{\milli\meter} and offset = \SI{150}{\milli\meter}.}
    \label{fig:longSlotReflectedWakepotential}
\end{figure}

\begin{figure}[h]
    \includegraphics[width=8cm]{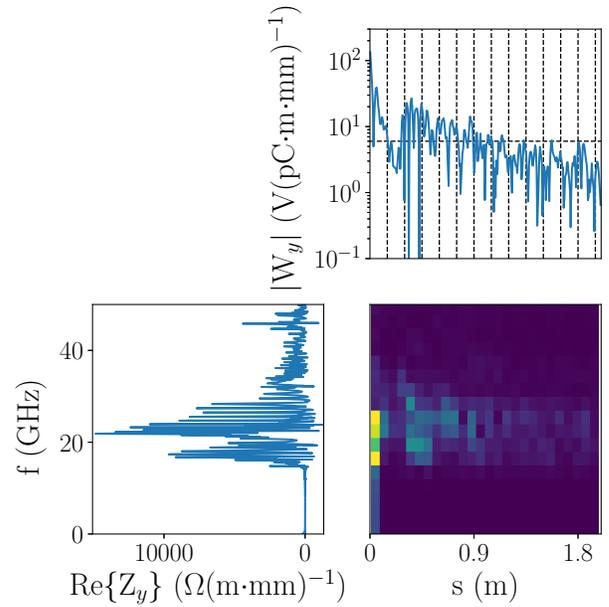}
    \caption{Evolution of the mode content of the reflected wakepotential with offset = \SI{100}{\milli\meter}.}
    \label{fig:spectrogram}
\end{figure}

An example of a reflected wakepotential is shown in Figure \ref{fig:longSlotReflectedWakepotential}. After the drive bunch passes through the structure, the remaining wakefields are extracted through the HOM waveguides. The fields propagating out of the cell through the coupler waveguide encounter the electric boundary and are reflected back into the cell. At this time, the wakepotential diverges from the open boundary case, increasing above the wakepotential threshold. The position where the reflected wakepotential diverges from the open boundary case is determined by the distance between the electric boundary and the beam axis. We call this distance the offset. The spectrogram of Figure \ref{fig:spectrogram} shows the mode evolution of the reflected wakepotential. The 17 and \SI{22}{\giga\hertz} modes of Figure \ref{fig:longSlotWakeImp} are quickly damped. The reflected wakepotential first consists of modes near \SI{17}{\giga\hertz}. While these modes are being damped, modes near \SI{22}{\giga\hertz} arrive. The length of the reflected wake is due to the difference in arrival time of the two sets of modes and because of the slow damping of these modes. Increasing the offset decreased the maximum amplitude of the reflected wakepotential. A larger offset accentuated the difference in arrival time causing the cell to damp one set of modes at a time. The reflected wakepotential duration increased as a result. The modes also spread in time due to dispersion in the waveguides.

The electric boundary, position of the distribution network relative to the beam, can be adjusted such that the reflected wake arrives after the a particular bunch but because the reflected wakepotential is damped slowly and lasts longer than the bunch spacing, the reflected wakepotential would be above the threshold at the next bunch. Increasing the HOM waveguide aperture and width, which previously had an effect on the wakepotential, proved ineffective at reducing the reflected wakepotential. The strategy of evaluating different iris geometries in an attempt to change the mode beating to form a minimum at the bunch located in the reflected wakepotential wasof unsuccessful.

\section{\label{sec:reflectedDesigns}Designs to reduce reflected wakefields}

\begin{figure}[h]
    \includegraphics[width=5cm]{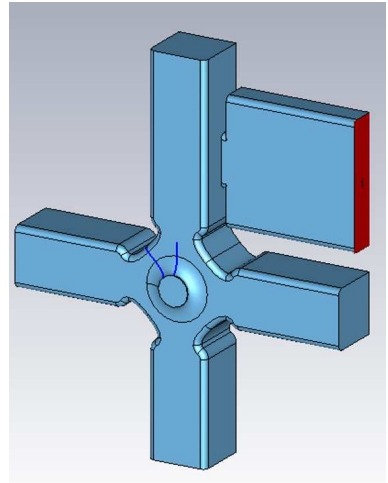}
    \caption{Horizontal coupler cell geometry.}
    \label{fig:horizontalCouplerGeometry}
\end{figure}

\begin{figure}[h]
    \includegraphics[width=8cm]{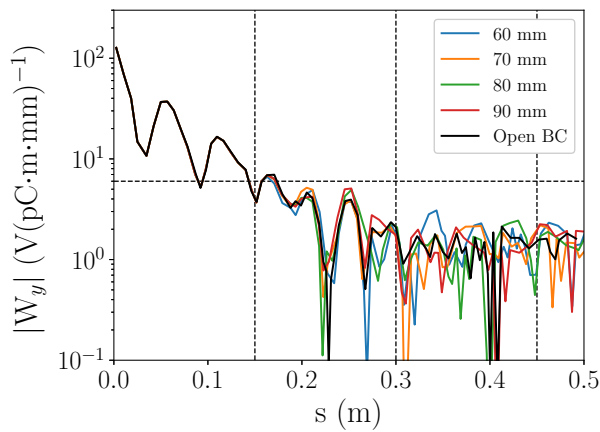}
    \caption{Reflected transverse wakepotential of horizontal coupler cell for various offsets.}
    \label{fig:horizontalCouplerWake}
\end{figure}

Two configurations were found to reduce the reflected wakepotentials to an acceptable level. First, the horizontal coupler cell design shown in Figure \ref{fig:horizontalCouplerGeometry} feeds power to the cell from a coupler waveguide through an aperture in one of the HOM waveguides. The aperture's dimensions and location allow power to enter the cell to establish the accelerating field while being sufficiently small that the wakefields couple weakly to the coupler waveguide. The wakefields are damped in the HOM waveguides instead of being reflected at the coupler waveguide. The reflected wakepotentials of Figure \ref{fig:horizontalCouplerWake} do not depend strongly on the offset meaning the horizontal coupler is transparent to the wakefields and a distribution network can be designed separately.

\begin{figure}[h]
    \includegraphics[width=3cm]{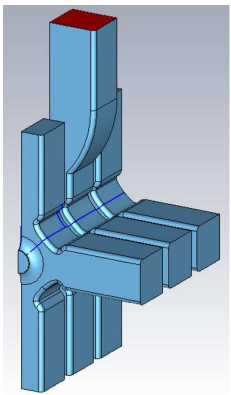}
    \caption{Triplet module based on the long slot cell with half symmetry.}
    \label{fig:tripletGeometry}
\end{figure}

\begin{figure}[!h]
    \includegraphics[width=8cm]{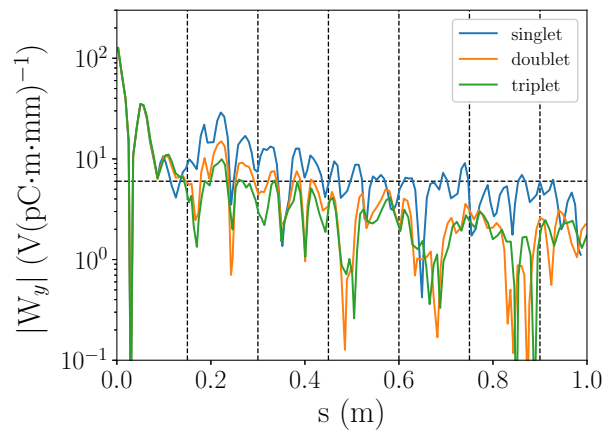}
    \caption{Reflected transverse wakepotentials of structures composed of singlets (long slot cell), doublet (one long slot and one four HOM waveguide cells) and triplet (one long slot and two four HOM waveguide cells).}
    \label{fig:tripletWake}
\end{figure}

A second scheme where modules were formed of coupler cells and four HOM waveguide cells like the triplet shown in Figure \ref{fig:tripletGeometry} also reduced the reflected wakepotentials below the CLIC threshold. The reflected wakepotential amplitudes were lowered by reducing the number of coupler waveguides from which reflections occur and by increasing the number of HOM waveguides that damp the wakefields. The module takes advantage of an iris size of \SI{2.75}{\milli\meter}, which is large enough to allow for coupling between cells. Operating a group of modules as a structure reduces the total number of coupler waveguides required compared to distributed coupling structure designs where each cell has a feed to a power source. Figure \ref{fig:tripletWake} shows the reflected wakepotential amplitude decreases as the number of four HOM waveguide cells increases. The wakepotential of the triplet is below the threshold at all bunch positions out to \SI{1}{\meter}. The outer radii of the cells in the doublet and triplets need to be tuned to achieve a flat longitudinal field. A triplet module where the coupler cell is a horizontal coupler cell, previously shown to considerably reduce the reflected wakepotential, combines a reduction of feed waveguides and effective mitigation of reflected wakefields.

\section{\label{sec:conc}Conclusion}
\vspace{1pt}

A SW cell applicable to distributed coupling structures was developed. The HOM waveguide width, w, was effective at lowering the wakepotential at the second bunch position. The wakepotentials were found to be generally less sensitive to changes in the HOM waveguide aperture, iw. From the SW cell, two configurations were developed capable of supporting \SI{100}{\mega\volt/\meter} loaded gradient and gave transverse wakepotentials below the threshold. The wakefields horizontal coupler cell couple weakly to the high power waveguide and are kept close to open boundary levels. The triplet module increases the number of HOM waveguides and decreases the number of coupler waveguides from which reflections originate to keep the reflected wakepotential low.

\bibliography{distributedCouplingWakes.bib}

\end{document}